\documentclass[12pt]{article}
\usepackage{amssymb,amsmath,graphicx,latexsym}\pdfoutput=1
\begin{document}
\title{\textbf{Dynamical symmetry breaking, CP violation, and a Higgs-like particle}} 
\author{B.~Holdom\thanks{bob.holdom@utoronto.ca}\\
\emph{\small Department of Physics, University of Toronto}\\[-1ex]
\emph{\small Toronto ON Canada M5S1A7}}
\date{}
\maketitle
\begin{abstract}
A pseudoscalar associated with dynamical symmetry breaking can mix with a heavier scalar when CP is violated. We study how such a mixed state that is mostly pseudoscalar can resemble the observed 125 GeV particle. The production rate from $gg$ is enhanced by the new heavy quarks and this compensates for the reduced $WW$, $ZZ$ and $\gamma\gamma$ branching ratios. The particle can appear to be a scalar from angular distributions in $ZZ$ decay while its pseudoscalar couplings should become evident in other processes.
\end{abstract}

Dynamical electroweak symmetry breaking suggests that the 125 GeV Higgs-like particle should have little to do with a field whose expectation value breaks the symmetry. The light bosons that are expected are pseudoscalar and such particles should have tiny branching ratios to $WW$ and $ZZ$. The data is suggesting otherwise \cite{:2012gk,:2012gu}. An implicit assumption behind the pure pseudoscalar prediction is CP invariance, and so it is now worthwhile to consider relaxing this. We shall focus on a strongly interacting fourth family as a specific framework for strongly interacting chiral fermions. This is similar to a one family technicolor model but the differences are more involved than a simple change of notation \cite{Holdom:2012pw}. Our discussion could be generalized to any set of condensing fermions of a general dynamical symmetry breaking theory.

We shall represent the dynamics at a TeV in terms of $SU(2)_L\times U(1)$ invariant 4-fermion operators involving the four heaviest fermions, $t'$, $b'$, $\tau'$ and $t$. We shall initially ignore the fourth neutrino, $\nu'$; we return to it at the end where we consider two possibilities for its mass, Dirac or Majorana. Some new strong gauge interaction is assumed to be broken close to a TeV and give rise to the following effective operators. 
\begin{equation}
\overline{q}'_Lt'_R\overline{t}'_Rq'_L\quad\quad\overline{q}'_Lb'_R\overline{b}'_Rq'_L\quad\quad\overline{\ell}'_L\tau'_R\overline{\tau}'_R\ell'_L\quad\quad\overline{q}_Lt_R\overline{t}_Rq_L
\label{e1}\end{equation}

Electroweak symmetry breaking can be driven by the condensates of $t'$ and $b'$, in which case the coefficients of at least the first two operators in (\ref{e1}) must be above some critical value (as in the NJL model \cite{nambu}). The critical value is somewhat reduced for the quark operators since QCD is also attractive in these channels. We emphasize that the modelling of gauge dynamics by these effective operators is not a controlled approximation, and it may not even provide the right explanation for the pattern of mass among these fermions (we say a little more about this at the end). But this caricature of the TeV dynamics is sufficient to explore the effects of CP violation.

For CP violation we need to consider another set of operators. The operators in (\ref{e2}) couple together the scalar bilinears composed of the $t'$, $b'$, $\tau'$ and $t$ flavours. (Four of these operators have a LRLR structure where the $SU(2)_L$ indices are contracted with $i\sigma_2$.)
\begin{eqnarray}
\overline{q}'_Lb'_R\overline{\tau}'_R\ell'_L\quad\quad\overline{q}'_Lt'_R\overline{q}'_Lb'_R\quad\quad\overline{q}'_Lt'_R\overline{\ell}'_L\tau'_R\nonumber\\
\overline{q}'_Lt'_R\overline{t}_Rq_L\quad\quad\overline{q}'_Lb'_R\overline{q}_Lt_R\quad\quad\overline{\ell}'_L\tau'_R\overline{q}_Lt_R
\label{e2}\end{eqnarray}
The operators of (\ref{e2}) are not hermitian (the hermitian conjugates are not shown) and so it is possible that they can carry CP violating phases. In fact the phases of these operators are sensitive to axial phase rotations of the four Dirac fermions.\footnote{Other operators may be sensitive to more general transformations and thus play a role in vacuum orientation, but they are not relevant for our discussion.} But one combination of these rotations is contained in $SU(2)_L\times U(1)$ and is thus inert. The remaining three phase freedoms cannot in general remove the six phases in the six operators. Any four of the operators would be sufficient to have a physical CP violating phase. If the three operators involving $t$ (or $\tau'$) were removed, again a physical phase would remain.

We assume that none of the operators of (\ref{e2}) can be generated by the TeV scale gauge interactions and that they have a different origin in some new flavour physics at a higher scale. Their coefficients are then suppressed relative to those in (\ref{e1}). But they need not be much smaller since anomalous scaling effects due to the strong gauge interaction can keep the operators close to being relevant. In fact one of the operators of (\ref{e2}) may be largely responsible for the top mass,\footnote{The second operator in the second row may be favoured \cite{Holdom:2006mr}.} in which case $m_t/m_{q'}$ gives an estimate of the size of an operator in (\ref{e2}) relative to the first two operators in (\ref{e1}).

We would like to move to a description that brings in the propagating bosonic degrees of freedom of interest. The standard approach involves the introduction of auxiliary complex scalar fields that allows the 4-fermion terms to be replaced by a sum of Yukawa terms and mass terms for the scalar fields. Fermion loops in this equivalent description generate gauged kinetic terms for the scalar fields. They also generate a scalar potential and negative mass terms that in this language drives the symmetry breaking. This construction in the context of the $t'$, $b'$, $t$ system was carried out in \cite{Hashimoto:2009ty}. We shall end up with an example of a multi-Higgs model, and the study of CP violation in such models, in particular the two-Higgs-doublet model, has a long history as reviewed in  \cite{Kraml,Branco:2011iw}. CP violation can occur explicitly or spontaneously with the latter case receiving attention in technicolor \cite{Lane:2000es}.

We define the complex scalar field doublets in a flavour basis $\phi_{t'}$, $\phi_{b'}$ , $\phi_{\tau'}$ and $\phi_{t}$. They all have hypercharge 1/2 and they are defined to couple to the fermions as follows ($\tilde{\phi}\equiv i\sigma_2\phi^*$),
\begin{equation}
-{\cal L}_{\rm Yukawa}= \overline{q}'_Lt'_R\tilde{\phi}_{t'}+ \overline{q}'_Lb'_R\phi_{b'}+ \overline{\ell}'_L\tau'_R\phi_{\tau'}+ \overline{q}_Lt_R\tilde{\phi}_{t}+h.c.
\end{equation}
The scalar mass terms can be expressed in terms of a hermitian matrix ${\cal M}^2$,
\begin{equation}
{\cal L}_{\rm mass}=-S^\dagger {\cal M}^2S,\quad\quad S^T\equiv(\phi_{t'},\phi_{b'},\phi_{\tau'},\phi_t) 
.\end{equation}
${\cal L}_{\rm mass}$ along with ${\cal L}_{\rm Yukawa}$ reproduces the original 4-fermion operators when ${\cal M}^2$ is the inverse of the matrix of coefficients of those operators.  The diagonal (off-diagonal) entries of the coefficient matrix are associated with the operators in (\ref{e1}) ((\ref{e2})). Both ${\cal M}^2$ and its inverse share the property that the off-diagonal elements are smaller than the diagonal elements. The diagonal elements of ${\cal M}^2$ receive contributions at one loop that drive at least some of these masses negative.  The off-diagonal elements of ${\cal M}^2$ carry the phases.

The kinetic terms for the scalar fields that are generated at one loop yield the $W$ and $Z$ masses when $\phi_i$ are replaced by $v_i$. The $W$ and $Z$ masses can also be obtained directly from the massive fermion loop and so
\begin{equation}
v_i^2=\frac{n_i m_i^2}{4\pi^2}\ln\frac{\Lambda_i}{m_i}
\label{e3}\end{equation}
where $n_i=3$ or 1 is the colour factor. $v_i^2$ is the contribution from this flavour to $v^2=(246\mbox{ GeV})^2$.  $\Lambda_i$ is the compositeness scale. $m_i$ is the fermion mass and it is related to $v_i$ via the Yukawa coupling $m_i=y_i v_i/\sqrt{2}$. By inserting this back into (\ref{e3}) we obtain the one loop values of the Yukawa couplings $y_i$. These couplings are the analog of the ratio of the constituent quark mass and $f_\pi$ in QCD.

A more realistic description of the underlying physics involves Schwinger-Dyson equations and momentum dependent dynamical fermion masses emerging as their solutions. The decrease of the fermion mass functions $\Sigma_i(p)$ for increasing $p$ naturally cuts off the one loop contributions and so the values of $m_i$ and $\Lambda_i$ serve to characterize this behaviour. The simplified local description should be adequate to show how CP violation can affect the neutral spinless degrees of freedom.

The higher powers of the fields in the scalar potential are also generated and at one loop are CP conserving. For example the quartic terms  in the case $y_{q'}=y_{t'}=y_{b'}$ take the simple form \cite{Hashimoto:2009ty}
\begin{eqnarray}
{\cal L}_4&=&y_{q'}^2[(\phi_{t'}^\dagger\phi_{t'})^2+(\phi_{b'}^\dagger\phi_{b'})^2+2(\phi_{t'}^\dagger\phi_{t'})(\phi_{b'}^\dagger\phi_{b'})-2(\phi_{t'}^\dagger\phi_{b'})(\phi_{b'}^\dagger\phi_{t'})]\nonumber\\&&+\;y_{\tau'}^2(\phi_{\tau'}^\dagger\phi_{\tau'})^2+y_{t}^2(\phi_{t}^\dagger\phi_{t})^2
.\end{eqnarray}
So at one loop we have the case of explicit CP violation originating only in the quadratic terms, i.e. `soft' breaking of CP. At higher loop order the more general set of terms in the potential is generated.

We have assumed that there is a charge conserving minimum of the potential at nonvanishing values of the neutral components of the fields $\langle\phi^0_{t'}\rangle$, $\langle\phi^0_{b'}\rangle$, $\langle\phi^0_{\tau'}\rangle$ and $\langle\phi^0_t\rangle$. All four vevs can be expected to be nonvanishing even if not all of the diagonal elements of ${\cal M}^2$ are negative, due to the off-diagonal elements of ${\cal M}^2$ \cite{Hashimoto:2009ty}. In  general the vevs are complex but the phase symmetry from $SU(2)_L\times U(1)$ and further field redefinitions can move these phases out of the vevs, thereby changing the phases in ${\cal M}^2$. (In the case of spontaneous CP violation the phases first appear in the vevs, but once the phases are rotated out of the vevs the situation is the same.) The phases of the fermions can also be adjusted to keep the Yukawa couplings real. Effectively the phases of the original 4-fermion operators are being adjusted (not removed) so that the fermion condensates are real.

The neutral components of the scalar fields can now be written in terms of the scalars $\sigma_i$ and pseudoscalars $\pi_i$ 
\begin{equation}
\phi^0_i=\frac{1}{\sqrt{2}}(v_i+\sigma_i)e^{i\epsilon_i\pi_i/v_i}
\end{equation}
where $\epsilon_{t'}=\epsilon_t=-1$, $\epsilon_{b'}=\epsilon_{\tau'}=1$. With this definition $(\pi_{t'}+\pi_{b'})/\sqrt{2}$ is an isosinglet. The off-diagonal terms in ${\cal M}^2$ are the source of the explicit breaking of the phase symmetries and so they provide masses for the $\pi_i$'s. One pseudoscalar remains massless, the Goldstone boson $\sum_i \epsilon_i v_i\pi_i/v$ absorbed by the $Z$.\footnote{In the context of technicolor another pseudoscalar is removed from the light spectrum due to technicolor instantons.} The expansion of off-diagonal terms with complex coefficients can also give rise to $\sigma_i\pi_j$ terms. This is the main physics of interest here, the mass mixing that can occur between the scalars $\sigma_i$ and the pseudoscalars $\pi_i$ due to CP violation.

The full mass matrix must be diagonalized to obtain the physical states. It is convenient to express these states in terms of the $\sigma_i$ and $\pi_i$ modes that are respectively CP even and odd, where for example the $\pi_i$ do not have tree level couplings to gauge bosons. The CP violation will show up as states that have mixed CP properties. Let us focus on a light mass eigenstate, labelled $\Pi$, to be associated with the 125 GeV resonance. We express it as
\begin{equation}
\Pi=\sum_i p_i\pi_i+\sum_i s_i\sigma_i
.\label{e4}\end{equation}
Since it is light this state should be dominated by its $\pi_i$ components. Also $\sum_i \epsilon_i v_i p_i=0$ since it must be orthogonal to the Goldstone mode. This allows it to have a large overlap with the isosinglet $(\pi_{t'}+\pi_{b'})/\sqrt{2}$, which is what we shall assume, and this ensures its coupling to $gg$. This particular isosinglet pseudoscalar was studied in \cite{Holdom:2012pw}. The charged pseudoscalars are contained in an isotriplet, but this should be dominated by its leptonic modes due to the required orthogonality to the isotriplet Goldstone modes. There are two other neutral pseudoscalars, one that may be dominated by the leptonic isosinglet mode and one that may be dominated by the top mode. Our focus shall be on $\Pi$.

We should now consider the mass matrix $M^2$ in the basis $(\pi_{t'},\pi_{b'},\pi_{\tau'},\pi_t,\sigma_{t'},\sigma_{b'},\sigma_{\tau'},\sigma_t)$. $M^2$ is real and symmetric and diagonalized with an orthogonal transformation. The real mixing parameters in (\ref{e4}) satisfy $\sum_i p_i^2+\sum_i s_i^2=1$ and they also appear in the inverse transformations, $\pi_i=p_i\Pi+...$, $\sigma_i=s_i\Pi+...$. These parameters thereby determine the $\Pi$ couplings to other fields.

Let us first consider the tree level $W$ and $Z$ couplings, which vanish for the $\pi_i$'s. Since the $\phi_i$ are standard doublets, each $\sigma_i$ has a coupling to the same combination of $WW$ and $ZZ$ as the standard Higgs, but with a coupling suppressed by $v_i/v$. Thus the width of the $\Pi$ into $WW$ or $ZZ$ is $v^{-2}(\sum_i s_i v_i)^2$ times the standard Higgs width into $WW$ or $ZZ$. We note that the combination $\sum_i v_i\sigma_i/v$ has the standard Higgs widths and thus corresponds to the standard Higgs. But here $\sum_i v_i\sigma_i/v$ is generally not a mass eigenstate and $\Pi$ can have an overlap with this combination, $(\sum_i s_i v_i)^2\neq0$. Effectively $\Pi$ needs to be mixed with a standard heavy Higgs to boost its decay to $WW$ and $ZZ$.

The $\Pi$ couplings to the heavy fermions are given by
\begin{equation}
\Pi\left(\sum_i\frac{m_i}{v_i} \overline{\psi}(s_i+ip_i\gamma_5)\psi_i\right)
.\end{equation}
From this we find that $\Gamma(\Pi\rightarrow gg)$ is enhanced relative to the standard Higgs by a factor
\begin{equation}
R_g=\frac{9}{4}\left(\sum_{i=t',b',t}\frac{p_i}{v_i}\right)^2v^2+\left(\sum_{i=t',b',t}\frac{s_i}{v_i}\right)^2v^2
.\end{equation}
We have used the infinite mass approximation for the quark loops. This result should still be a good approximation after QCD corrections, since the corrections to scalar and pseudoscalar $gg$ couplings are very similar at NNLO \cite{Harlander:2002vv,Anastasiou:2002wq}.\footnote{This fact was missed in \cite{Holdom:2012pw}.} For the pseudoscalar combination $(v_{b'}\pi_{t'}+v_{t'}\pi_{b'})/\sqrt{v_{t'}^2+v_{b'}^2}$ the resulting $R_g=(9/4)v^2(v_{t'}^2+v_{b'}^2)/(v_{t'}^2v_{b'}^2)$ is greater than 9.  Thus we expect the $gg$ production mode of the $\Pi$ to be significantly enhanced relative to a standard Higgs. In order to obtain the observed production of $WW$ and $ZZ$, the branching ratio of $\Pi$ into $WW$ and $ZZ$ must be reduced compared to the standard Higgs. This happens naturally here because of the mixing suppression factor $v^{-2}(\sum_i s_i v_i)^2$.

The enhancement of the $gg$ width by new heavy fermions is not seen in standard CP violating two-Higgs-doublet models. Without this enhancement, the light mixed scalar-pseudoscalar state in such models tends to be pushed to the pure scalar limit in order to recover sufficient $ZZ$ and $WW$ production \cite{Barroso:2012wz,Freitas:2012kw}.

A suppression factor $v^{-2}(\sum_i s_i v_i)^2\sim1/10$ is needed to produce $WW$ and $ZZ$ production rates similar to that of the Higgs. The suppression factor resembles the square of a mixing angle and so it might be expected to be comparable to the ratio between the smallest and largest eigenvalues of the matrix $M^2$. The latter is $\sim1/100$ if the lightest of the boson states is 125 GeV and the heaviest is around a TeV. But since $M^2$ has only positive eigenvalues (a stable minimum), the square root of these values are the eigenvalues of another symmetric matrix (the square root of $M^2$) that is diagonalized by the same orthogonal transformation that diagonalizes $M^2$. This suggests that a suppression factor of $\sim1/10$ is not unreasonable for the given mass ratios, and a study of random choices for $M^2$ bears this out.

$\Gamma(\Pi\rightarrow \gamma\gamma)$ is reduced relative to the standard Higgs with a suppression factor
\begin{equation}
R_\gamma=\frac{\frac{9}{4}\left[\sum_ip_i\frac{n_i q_i^2v}{v_i}\right]^2+\left[\sum_is_i(\frac{v_i}{v}C_W-\frac{n_i q_i^2v}{v_i})\right]^2}
{\left(C_W-\frac{4}{3}\right)^2}
.\end{equation}
$n_i$ is the colour factor, $q_i$ is the charge and $C_W\approx6.26$ due to the $W$ loop. The pseudoscalar piece is reduced because of the absence of the $W$ while the scalar piece is reduced because more fermions contribute. We give $R_\gamma$ for two special cases; $R_\gamma\approx0.26$ for the pseudoscalar with $p_{t'}=p_{b'}=v_{t'}/v=v_{b'}/v=1/\sqrt{2}$ and $R_\gamma\approx0.21$ for the Higgs scalar combination. From these examples we see that $R_gR_\gamma\sim2$ which is an enhancement factor in the production rate of $\gamma\gamma$ relative to a standard Higgs. As for the total width of $\Pi$, the increase in $\Gamma(\Pi\rightarrow gg)$ more than compensates for the decrease in $\Gamma(\Pi\rightarrow WW,ZZ)$, but $\Gamma(\Pi\rightarrow b\overline{b},c\overline{c},\tau\overline{\tau})$ may also decrease with the result that the total width of $\Pi$ can be similar to the Higgs.

A first experimental test of this picture will be to see the reduced $W$ and $Z$ couplings of $\Pi$. The 125 GeV particle should have smaller than expected production through vector boson fusion or when associated with $W/Z$. Further tests will need to probe for the existence of the mixed pseudoscalar-scalar couplings. In particular the $\Pi$ coupling to $ZZ$ is essentially scalar, that is it is dominated by the tree level couplings due to the $\sigma_i$ components of $\Pi$, and so the angular distributions of the four leptons from the $ZZ$ decay will be close to the Higgs result. This is line with the first results \cite{:2012br}.

The couplings of $\Pi$ to $\gamma\gamma$ and $gg$ are definitely mostly pseudoscalar. The couplings to $t$, $b$ and $\tau$ will also have pseudoscalar components. Thus the main experimental objective would be to find evidence for any of these pseudoscalar couplings. In particular the pseudoscalar $gg$ coupling could be detected in the azimuthal angular separation distribution of the two jets in $\Pi+\mbox{2 jets}$ production \cite{Klamke:2007cu}.

The $\Pi$ coupling to fermions such as $t$, $b$ and $\tau$ should be at least roughly proportional to mass. But other than that the couplings are quite uncertain. The $t$ couplings depend on the mixing parameters $p_t$ and $s_t$ and the couplings to $b$ and $\tau$ are further complicated by the physics responsible for their mass. The relevant 4-fermion operators can bring in additional phases, and more than one operator can contribute to a given mass. But since $\Pi$ is predominantly pseudoscalar, its couplings to any of these fermions should continue to have significant pseudoscalar components. This could show up in the decay distributions of $\tau^+\tau^-$ if $BR(\Pi\rightarrow\tau^+\tau^-)$ is not reduced too much. Another possibility is $t\overline{t}$ associated production with, for example, $\Pi\rightarrow b\overline{b}$. The literature on these and other searches for CP violation in enlarged Higgs sectors is reviewed in \cite{Kraml}.

Thus far we have ignored the presence of the fourth neutrino $\nu'$. If it has a Dirac condensate then it can be treated similarly to the $\tau'$, and there would be a fifth scalar doublet $\phi_{\nu'}$. There would be four more operators with possible CP violating phases for a total of $10-4$ physical phases.

More interesting is the case of a Majorana neutrino mass. This is the case if the field $\nu'_R$ doesn't exist in the TeV scale theory, leaving the purely Majorana condensate $\langle\nu'_L\nu'_L\rangle$ as the only possibility. This leads to the introduction of a $SU(2)_L$ triplet scalar field with a Yukawa term
\begin{equation}
\ell_L'^T\Phi_{\nu'}\ell_L'\mbox{ with }\Phi_{\nu'}\equiv\left(\begin{array}{cc}\Phi^0 & \frac{\Phi^+}{\sqrt{2}} \\\frac{\Phi^+}{\sqrt{2}} & \Phi^{++}\end{array}\right)
.\end{equation}
We set
\begin{equation}
\Phi_{\nu'}^0=\frac{1}{\sqrt{2}}(v_{\nu'}+\sigma_{\nu'})e^{-i\pi_{\nu'}/v_{\nu'}}
.\end{equation}
Fermion loops containing the neutrino contribute to the $W$ and $Z$ masses and the $\log\Lambda$ terms are captured by the kinetic terms of the triplet scalar \cite{Gelmini:1980re}. This gives
\begin{eqnarray}
m_W^2&=&\frac{g^2}{4}(v_{t'}^2+v_{b'}^2+v_{\tau'}^2+v_{\nu'}^2)\\
m_Z^2&=&\frac{g^2+g'^2}{4}(v_{t'}^2+v_{b'}^2+v_{\tau'}^2+2v_{\nu'}^2)
\end{eqnarray}
where
\begin{equation}
v_{\nu'}^2=\frac{m_{\nu'}^2}{4\pi^2}\ln\frac{\Lambda_{\nu'}}{m_{\nu'}}
.\end{equation}

The extra contribution to $m_Z^2$ means a negative contribution to the $T$ parameter $\alpha\Delta T=-v_{\nu'}^2/v^2$ due to the Majorana neutrino mass \cite{Holdom:1996bn}. For example $\Delta T\approx-0.4$ for $m_{\nu'}=100$ GeV and $\Lambda_{\nu'}=2m_{\nu'}$. There are also contributions to the $W$ and $Z$ masses from the underlying physics that are finite (independent of $\Lambda$), in particular the positive contributions to $T$ arising from fermion mass splittings. The new negative contribution is then welcome, since it allows more $t'-b'$ and $\nu'-\tau'$ mass splitting. $m_{\tau'}>m_{\nu'}$ is especially favoured since the contribution to $S$ from leptons is
\begin{equation}
S_{\rm leptons}\approx \frac{1}{6\pi}-\frac{1}{3\pi}\ln(\frac{m_{\tau'}}{m_{\nu'}})-\frac{1}{12\pi}
,\end{equation}
where the last term is another consequence of the Majorana mass. $S_{\rm leptons}$ is allowed to become more negative and the viability of the fourth family is improved \cite{Holdom:2006mr}. Most studies of the fourth family ignore this ``pure Majorana'' possibility.

$SU(2)_L\times U(1)$ invariant operators that couple $\nu'_L\nu'_L$ to other fermions have 6 fermions rather than 4. Thus cubic terms appear in the scalar potential, such as $\tilde{\phi}^T_{t'}\Phi_{\nu'}\tilde{\phi}_{t'}$ and $\phi_{b'}^T\sigma_2\Phi_{\nu'}^*\sigma_2\phi_{b'}$. Such terms can be another source of CP violation and in this case could cause $\Pi$ to pick up a $\sigma_{\nu'}$ component. The interest here is that the $\sigma_{\nu'}$ coupling to $ZZ$ versus $WW$ is twice as large as for a standard Higgs. $\Gamma(\Pi\rightarrow ZZ)/\Gamma(\Pi\rightarrow WW)$ could then deviate from standard expectations. But the $\pi_i\sigma_{\nu'}$ mixing terms are expected to be smaller than before because 6-fermion operators should be smaller than 4-fermion operators. So unless the $\sigma_{\nu'}$ mass before mixing was unexpectedly close to the $\Pi$ mass the size of the effect appears to be small.

In multi-Higgs models light scalars are typically the main attraction, but here they are relegated to large masses, both neutral and charged. Among the neutral scalar fields before mass mixing, $\sigma_{t'}$ and $\sigma_{b'}$ are related to 4-fermion operators that are driving the symmetry breaking. These masses at one loop are typically close to $2m_i$ \cite{Bardeen:1989ds}. The only way for the $\sigma_i$'s and fermions to be light compared to the compositeness scale is through a fine tuning which can be traced back to the need for the original 4-fermion operators to be tuned close to critical strength. Since we don't assume fine tuning we expect the $\sigma_i$ to have TeV scale masses. For consistency the masses should be below the compositeness scales and we note that in QCD the sigma resonance at 441 MeV \cite{Caprini:2005zr} is markedly below $2m$ as well. Scaling up the QCD numbers gives a very broad scalar resonance around $1.2$ TeV and heavy quark masses around 750 or 800 GeV.

As for the top mass our discussion has suggested that the last operator in (\ref{e1}) is subcritical, to ensure that the top mass is smaller than the $t'$ mass. But here we note that this is not strictly necessary. $v_{t'}\gg v_t$ (and thus $m_{t'}\gg m_t$) could be a property of a minimum of the full potential even if the potential was roughly symmetric between $t'$ and $t$. To spontaneously break such an approximate discrete symmetry, the potential would just need to contain a sufficiently large term of the form $(\phi_{t'}^\dagger\phi_{t'})(\phi_{t}^\dagger\phi_{t})$. An example of a TeV scale broken gauge interaction that treats $t'$ and $t$ similarly is a $U(1)$ under which the third and fourth families have equal and opposite charges.\footnote{See \cite{Holdom:2008xx,Holdom:2011fc} for some discussion of such a $U(1)$. Also a strong $U(1)$ gauge theory with many flavours has nontrivial characteristics \cite{Holdom:2010qs}.}

If the origin of CP violation lies above the weak scale then it could be arising spontaneously along with the breaking of flavour gauge symmetries. This flavour physics may be in the 100 to 1000 TeV range. It would be quite remarkable if the first particle discovered at the LHC provides a window onto this physics. We have been fairly qualitative in our discussion because there is presently little handle on the sizes of the various operators of interest.  The exception are those operators responsible for the top mass, and they support the notion that the operators of interest are not small. Despite the uncertainties we have ended up with some properties for a light boson that are very distinctive, and probably distinctive enough to be tested in the current LHC run. More theoretical questions of how this picture could fit in with consistent descriptions of weak and strong CP violation can await these tests.

\section*{Acknowledgments}
This work was supported in part by the Natural Science and Engineering Research Council of Canada.


\begin{thebibliography}{99}
\bibitem{:2012gk} 
  G.~Aad {\it et al.}  [ATLAS Collaboration],
  Phys.\ Lett.\ B {\bf 716}, 1 (2012)
  [arXiv:1207.7214 [hep-ex]].
\bibitem{:2012gu} 
  S.~Chatrchyan {\it et al.}  [CMS Collaboration],
  Phys.\ Lett.\ B {\bf 716}, 30 (2012)
  [arXiv:1207.7235 [hep-ex]].
\bibitem{Holdom:2012pw} 
  B.~Holdom,
  Phys.\ Lett.\ B {\bf 709}, 381 (2012)
  [arXiv:1201.0185 [hep-ph]].
\bibitem{nambu} Y.~Nambu and G.~Jona-Lasinio, Physical Review 122: 345 (1961).
    \bibitem{Holdom:2006mr} 
  B.~Holdom,
  JHEP {\bf 0608}, 076 (2006)
  [hep-ph/0606146].
\bibitem{Hashimoto:2009ty} 
  M.~Hashimoto and V.~A.~Miransky,
  Phys.\ Rev.\ D {\bf 81}, 055014 (2010)
  [arXiv:0912.4453 [hep-ph]].
  \bibitem{Kraml}
S. Kraml et al., Chapter 2 in CERN-2006-009, hep-ph/0608079.
\bibitem{Branco:2011iw} 
  G.~C.~Branco, P.~M.~Ferreira, L.~Lavoura, M.~N.~Rebelo, M.~Sher and J.~P.~Silva,
  Phys.\ Rept.\  {\bf 516}, 1 (2012)
  [arXiv:1106.0034 [hep-ph]].
  \bibitem{Lane:2000es} 
  K.~D.~Lane, T.~Rador and E.~Eichten,
  Phys.\ Rev.\ D {\bf 62}, 015005 (2000)
  [hep-ph/0001056].
    \bibitem{Harlander:2002vv} 
  R.~V.~Harlander and W.~B.~Kilgore,
  JHEP {\bf 0210}, 017 (2002)
  [hep-ph/0208096].
\bibitem{Anastasiou:2002wq} 
  C.~Anastasiou and K.~Melnikov,
  Phys.\ Rev.\ D {\bf 67}, 037501 (2003)
  [hep-ph/0208115].
    \bibitem{Barroso:2012wz} 
  A.~Barroso, P.~M.~Ferreira, R.~Santos and J.~P.~Silva,
  Phys.\ Rev.\ D {\bf 86}, 015022 (2012)
  [arXiv:1205.4247 [hep-ph]].
  \bibitem{Freitas:2012kw} 
  A.~Freitas and P.~Schwaller,
  arXiv:1211.1980 [hep-ph].
 \bibitem{:2012br} 
  S.~Chatrchyan {\it et al.}  [CMS Collaboration],
  arXiv:1212.6639 [hep-ex].
\bibitem{Klamke:2007cu} 
  G.~Klamke and D.~Zeppenfeld,
  JHEP {\bf 0704}, 052 (2007)
  [hep-ph/0703202 [HEP-PH]].
   \bibitem{Gelmini:1980re} 
  G.~B.~Gelmini and M.~Roncadelli,
  Phys.\ Lett.\ B {\bf 99}, 411 (1981).
 \bibitem{Holdom:1996bn} 
  B.~Holdom,
  Phys.\ Rev.\ D {\bf 54}, 721 (1996)
  [hep-ph/9602248].
   \bibitem{Bardeen:1989ds} 
  W.~A.~Bardeen, C.~T.~Hill and M.~Lindner,
  Phys.\ Rev.\ D {\bf 41}, 1647 (1990).
 \bibitem{Caprini:2005zr} 
  I.~Caprini, G.~Colangelo and H.~Leutwyler,
  Phys.\ Rev.\ Lett.\  {\bf 96}, 132001 (2006)
  [hep-ph/0512364].
  \bibitem{Holdom:2008xx} 
  B.~Holdom,
  Phys.\ Lett.\ B {\bf 666}, 77 (2008)
  [arXiv:0805.1965 [hep-ph]].
\bibitem{Holdom:2011fc} 
  B.~Holdom,
  Phys.\ Lett.\ B {\bf 703}, 576 (2011)
  [arXiv:1107.3167 [hep-ph]].
 \bibitem{Holdom:2010qs} 
  B.~Holdom,
  Phys.\ Lett.\ B {\bf 694}, 74 (2010)
  [arXiv:1006.2119 [hep-ph]].
\end{thebibliography}
\end{document}